\documentclass[sigconf, nonacm]{acmart}
\setcopyright{none} 
\renewcommand\footnotetextcopyrightpermission[1]{}
\setcopyright{none}               
\renewcommand\footnotetextcopyrightpermission[1]{} 

\usepackage{tabularx}
\usepackage{booktabs}   
\usepackage{makecell}   
\usepackage[utf8]{inputenc}
\AtBeginDocument{%
  }

\begin{document}

\title{Co-Refine: AI-Powered Tool Supporting Qualitative Analysis}

\author{Athikash Jeyaganthan}
\email{psyaj9@nottingham.ac.uk}
\affiliation{%
  \institution{University of Nottingham}
  \city{Nottingham}
  \country{United Kingdom}
}
\author{Kai Xu}
\email{kai.xu@nottingham.ac.uk}
\affiliation{%
  \institution{University of Nottingham}
  \city{Nottingham}
  \country{United Kingdom}
}
\author{Franziska Becker}
\email{franziska.becker@vis.uni-stuttgart.de}
\affiliation{%
  \institution{University of Stuttgart}
  \city{Stuttgart}
  \country{Germany}
}
\author{Steffen Koch}
\email{Steffen.Koch@vis.uni-stuttgart.de}
\affiliation{%
  \institution{University of Stuttgart}
  \city{Stuttgart}
  \country{Germany}
}

\renewcommand{\shortauthors}{Jeyaganthan et al.}

\begin{abstract}
Qualitative coding relies on a researcher’s application of codes to textual data. As coding proceeds across large datasets, interpretations of codes often shift (temporal drift), reducing the credibility of the analysis. Existing Computer-Assisted Qualitative Data Analysis (CAQDAS) tools such as NVivo and ATLAS.ti provide excellent support for data management and codebook organisation but offer no workflow for real-time detection of these drifts. We present Co-Refine, an AI-augmented qualitative coding platform that delivers continuous, grounded feedback on coding consistency without disrupting the researcher’s workflow. The system employs a three-stage audit pipeline that operates on every coding decision. Stage 1 computes deterministic embedding-based metrics that provide mathematical consistency signals. Stage 2 grounds LLM verdicts within $\pm 0.15$ of the deterministic scores. Stage 3 produces code definitions from previous coding patterns, creating a feedback loop that deepens the system’s understanding of each code. 

Requirements were elicited through semi-structured interviews with three qualitative researchers (U1--U3). The implemented system was evaluated through unit, integration, and end-to-end testing, supplemented by a formative user study. Co-Refine demonstrates that deterministic scoring can effectively constrain LLM outputs to produce reliable, real-time audit signals for qualitative analysis.

\textbf{Code and data availability:} The source code, LLM prompts, evaluation datasets, and supplementary materials (including the full SUS questionnaire and thematic analysis codes) are available upon reasonable request from the author.
\end{abstract}


\begin{teaserfigure}
  \includegraphics[width=\textwidth]{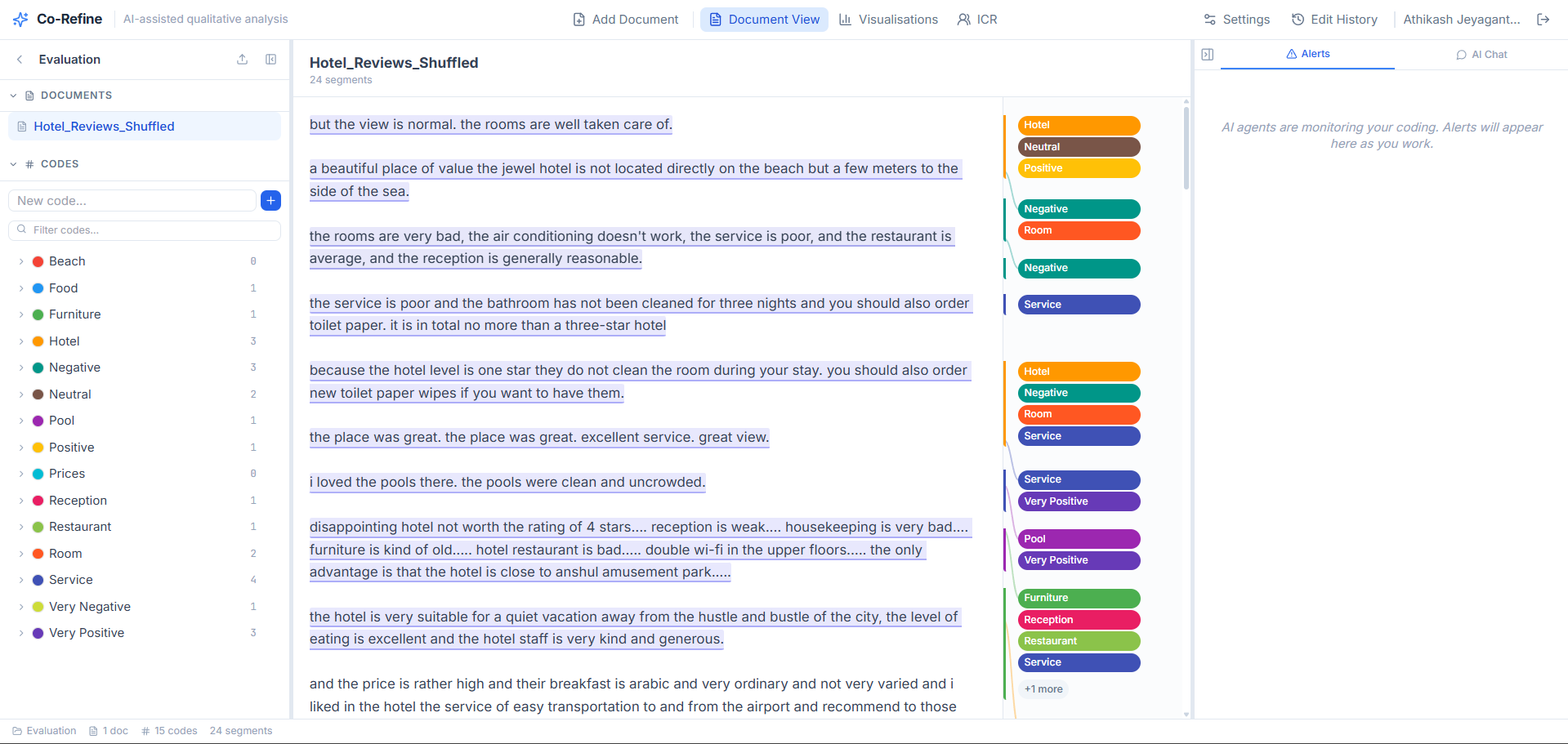}
  \caption{Overview of Co-Refine application.}
  \label{fig:doc-viewer-title}
\end{teaserfigure}

\maketitle

\section{Introduction}
\label{sec:intro}

Qualitative research remains central to the social sciences, humanities, and many applied domains. Researchers analyse interviews, open-ended survey responses, and documents to uncover meaning, patterns, and lived experiences. A foundational step in this process is \emph{qualitative coding} (QC), in which researchers assign conceptual labels — ``codes'' — to segments of unstructured text to identify themes and behaviours \cite{alma9923982638802466}. 

Yet coding is inherently challenging. Codes are not static; they evolve as researchers deepen their understanding of the data. This interpretive flexibility is a strength of qualitative work, but it also introduces \emph{temporal drift}: a researcher’s application of the same code can shift subtly over time, so that early and later instances no longer share the same meaning. Codes may also semantically overlap or become poorly differentiated, especially in large projects or collaborative settings where multiple coders must maintain a shared framework and demonstrate acceptable inter-coder reliability (ICR) \cite{cole2024interrater}. Without mechanisms to surface these inconsistencies in real time, researchers rely on memory and periodic manual reviews — a process that quickly becomes cognitively overwhelming and scales poorly as datasets grow.

The consequences are significant. Inconsistencies undermine the credibility of findings and make it difficult for peer reviewers to evaluate the reliability of emerging themes. One qualitative researcher we interviewed (U1) described a six-month setback when definitional drift was discovered only during the collaboration phase, forcing a complete re-examination of already-coded material. Another (U3) noted that the time required for consistency checks simply was not ``worth it,'' leading many to accept potential drift as inevitable. Existing Computer-Assisted Qualitative Data Analysis Software (CAQDAS) tools, such as NVivo and ATLAS.ti, help organise data and codebooks, but they offer no built-in support for detecting drift or semantic overlap during active coding. Researchers are left to perform these checks manually or at the end of analysis, often too late to act effectively.

Recent advances in machine learning and natural language processing present new opportunities. Embedding models can represent semantic relationships among text segments, while large language models (LLMs) can reason about intent and generate reflective feedback \cite{Siiman2023OpportunitiesAC}. However, most existing AI-assisted tools either fully automate coding (risking loss of researcher agency) or provide post-hoc analysis rather than continuous, actionable support within the researcher’s natural workflow.

In this paper, we present \textbf{Co-Refine}, an AI-augmented qualitative coding platform designed to maintain coding consistency as the codebook evolves---without automating the interpretive act itself. Co-Refine embeds a novel three-stage audit pipeline that is triggered at every coding decision. The first stage computes deterministic embedding-based metrics (centroid similarity and temporal drift) that provide transparent mathematical signals. The second stage grounds LLM-generated feedback within ±0.15 of these deterministic scores, producing plain-language findings and actionable suggestions tailored to qualitative researchers. A third stage periodically synthesises an evolving definition of each code from the researcher’s own coding patterns, creating a deepening feedback loop.

Our overall aim is to design and implement a reflective assistant that reveals inconsistencies, highlights semantic overlaps, and supports code refinement while preserving full researcher control at every step. By combining semantic similarity metrics, clustering techniques, and constrained LLM reasoning, Co-Refine enables researchers to engage more deeply with their data rather than delegating interpretation to the machine. In doing so, the system addresses a long-standing methodological gap in qualitative analysis: the lack of real-time, trustworthy consistency support that respects the interpretive, iterative, and human-centred nature of the work.

The remainder of this paper describes the design rationale and implementation of Co-Refine, presents results from automated pipeline evaluation and a formative user study, and discusses implications for future HCI tools that support qualitative research.

\section{Background and Related Works}
\label{sec:background}
Computer-Assisted Qualitative Data Analysis Software (CAQDAS) tools have become the industry standard for qualitative researchers. These solutions provide support for core stages of the qualitative analysis workflow. This includes importing and organising data in various formats, constructing hierarchical codebooks, applying codes to text segments, writing annotations, and retrieving coded segments via filtering mechanisms \cite{gao2024collabcoder, rietz2021cody}.

Visualisations include frequency charts, co-occurrence matrices, and concept maps. In addition, they support collaborative features that enable multiple users to work on the same project and perform ICR calculations, such as Cohen's kappa. These traits have enabled qualitative researchers to manage large datasets, thereby providing environments for systematic data analysis and theme development \cite{gebreegziabher2023patat, zade2018conceptualizing}.

Despite their value, existing CAQDAS tools are limited in the support they provide. Features such as consistency checking, semantic drift detection, and code overlap identification are not implemented; they are addressed only through manual reviews that reflect on previous code usage. This means that researchers must inspect their coded data periodically to ensure consistency, a process that scales poorly for codes exceeding a manageable limit \cite{rietz2021cody, gao2024collabcoder}.

Temporal drift and semantic code overlap go unnoticed unless explicitly recorded, leaving the integrity of the codebook to depend on the researcher's discipline to be wary of such issues. Researchers must periodically review all of their prior coded data, which becomes impractical as the depth of analysis increases. Currently, to address such issues, researchers must conduct peer reviews or sessions to discuss disagreements. At this point, it may already be too late; the drift may have already affected the coded data. Current CAQDAS workflows ensure that consistency is checked only at the end, rather than continuously, thereby missing the window when issues can be mitigated.

Recent studies in HCI have explored the use of supervised and unsupervised ML techniques to aid in qualitative coding. Early solutions incorporated rule-based pattern matching with supervised classifiers to propagate codes to unseen text. This offers a means to semi-automate the process while preserving human control through interactive rule editing \cite{rietz2021cody}. More complex findings have shown that leveraging neural embeddings and interactive program synthesis can learn symbolic rule sets across different features, generating suggestions that researchers can use \cite{gebreegziabher2023patat}.

Clustering embedding spaces has enabled researchers to discover concepts and thematic facets within coded datasets. LLooM combines embedded-based groupings with human-guided refinement to produce interpretable abstractions that mirror qualitative coding practices \cite{lam2024concept}. LLoom introduces a process known as "concept induction", which utilises four operators (Distill, Cluster, Synthesis, and Loop) to produce reliable interpretations from raw data. These methodologies, which employ ML techniques, significantly improve coding and ICR speed compared to manual workflows; however, they do not provide real-time feedback \cite{rietz2021cody, gebreegziabher2023patat, lam2024concept}.

LLMs have transformed the applications of AI for qualitative analysis. Tools such as CollabCoder integrate GPT models across their workflow. This includes generating code suggestions during independent coding, resolving disagreements, and supporting the finalisation of the codebook \cite{gao2024collabcoder}. CoAIcoder similarly utilises AI, serving as a shared mediator to accelerate collaborative decision-making while recording its decision history \cite{gao2023coaicode}.

LATA (LLM-Assisted Thematic analysis) has demonstrated that GPT-4 can achieve acceptable agreement with researchers, yielding a Cohen's Kappa of 0.73 \cite{wang2025lata}. In addition, specialised frameworks such as AbductiveAI integrate reasoning through chain-of-thought (CoT), thereby deepening the interpretive partnership. Many specialised systems have been developed, underlining the various applications of LLMs to qualitative research. 

DeTAILS \cite{sharma2025details} is a framework that maps LLM assistance directly to the thematic analysis framework theorised by Braun and Clarke through interactive feedback loops, which are designed to preserve researcher oversight. An editable codebook is used as an object so that any changes to codes or themes are automatically propagated through the coded data. 

Another solution that integrates LLMs is TIQA \cite{tseng2025tiqa}, which addresses the "burden" of research by applying design principles to reduce the distress researchers experience when analysing sensitive data. TIQA uses ML modules (SegmentModeler and CodeModeler) to identify content that may be perceived as "traumatic" and to provide warnings tailored to the user.

LOGOS \cite{pi2026logosllmdrivenendtoendgrounded} contributes a fully automated framework for the grounded theory workflow through semantic clustering and iterative refinement. A conceptual group is produced from the qualitative codes, which are aware of the semantic relationships among them. These relationships are classified as subsumption, equivalence, and orthogonality to create a theoretical framework. 

LLMs have also been used as proxies for participants or as assistants for analysis, raising significant questions about whether AI can represent real people and about potential bias in replacing human voices with simulations \cite{kapania2025simulacrum}. While solutions that use LLMs make analysis less demanding, they provide feedback after the fact rather than continuous, grounded audit signals during the coding process. Furthermore, researchers have raised concerns regarding the potential over-reliance on LLM outputs, as well as the risk of hallucination and overall loss of reflexive depth when there is a lack of constraint on outputs by deterministic evidence \cite{gao2024collabcoder, gao2023coaicode, kapania2025simulacrum}.  

Three limitations have been identified with existing systems:
\begin{itemize}
      \item AI assistance in current solutions is limited to post-hoc evaluation or conversational refinement rather than providing real-time deterministic and semantic scoring grounded in a user's own codebook.
    \item The integration of this into tools is absent from current solutions, so users are unaware of potential drifts during active coding.
    \item While existing collaborative tools may provide some ICR metrics to users, they have yet to implement useful conflict resolution workflows complemented by visualisations.
\end{itemize}
  
Overall, features that existing platforms have not yet implemented include deterministic centroid similarity, temporal drift quantification, and automatic facet clustering using KMeans and t-SNE.

From the identified limitations, a critical improvement to the current qualitative workflow is needed: a qualitative coding environment that provides continuous consistency feedback embedded within the researcher's workflow. This will require implementing several tools. First, a scoring method to quantify how well each new coding decision compares with the researcher's coding pattern for that code. Second, a warning for temporal drift that detects when the initial meaning of a code has shifted over time. Third, a constrained LLM feedback layer capable of generating feedback without overriding mathematical evidence and keeping hallucinations to a minimum. Fourth, a collaborative workflow that allows teams to identify and resolve coding disagreements. Addressing these gaps will make a valuable contribution to AI-assisted qualitative research.

\section{Co-Refine: Design and Implementation}
\label{sec:design}

We designed and implemented Co-Refine through a user-centered, iterative process that combined requirements elicitation with rapid prototyping and evaluation. The goal was to create a reflective assistant that provides real-time consistency feedback while preserving full researcher agency.

\subsection{Requirements Elicitation}
To ground the system in real qualitative research practice, we conducted semi-structured interviews with three experienced qualitative researchers (U1--U3) affiliated with UK institutions (3--12 years of experience). Interviews lasted 45--60 minutes and were guided by 11 open-ended questions covering data types, current coding workflows, tools, collaboration challenges, inter-coder reliability (ICR) practices, and attitudes toward AI assistance.

Participants relied primarily on non-CAQDAS tools (Excel, Miro) because existing platforms such as NVivo felt cumbersome. Major pain points included definition drift (U1 described a six-month setback discovered only during collaboration), time-intensive manual consistency checks (U3 noted the effort was ``not worth it''), and the lack of real-time support for semantic overlap or ICR. All participants expressed cautious openness to AI but stressed that trust, explainability, and human override were non-negotiable: AI must act ``as an assistant'' (U2) and be grounded in the researcher’s own data and coding history.

From these interviews we derived five core requirements that shaped every design decision:
\begin{itemize}
    \item Deterministic evidence must ground all LLM outputs to ensure trustworthiness.
    \item Human override must be preserved at every stage; the system is a reflective partner, not an autonomous coder.
    \item Researchers must be able to configure audit sensitivity and view deterministic scores when desired.
    \item The system must support multi-document projects and small-team collaboration with user-scoped data.
    \item Feedback must be plain-language, actionable, and integrated into the natural coding workflow.
\end{itemize}

These requirements directly informed the hybrid deterministic-LLM pipeline and the emphasis on researcher control.

\subsection{Design Principles}
Co-Refine was guided by three high-level principles:

\paragraph{Scalability} The system must handle arbitrarily long documents and growing numbers of coded segments without introducing latency. All audit computation runs in background threads; per-user ChromaDB collections and a persistent HNSW index ensure fast nearest-neighbour retrieval regardless of dataset size.

\paragraph{Modularity} Backend features are implemented as vertical slices (each containing its own router, service, repository, and schemas), while the frontend follows a four-layer architecture (\texttt{app/}, \texttt{widgets/}, \texttt{features/}, \texttt{shared/}). This structure kept the codebase maintainable as new capabilities (chat, visualisations, ICR) were added.

\paragraph{Usability and Researcher Control} AI agents serve only as advisors. Every alert includes a headline, plain-language finding, drift warning (if present), and actionable suggestion. Researchers can inspect deterministic scores, override any recommendation, configure project-level thresholds, and review an immutable edit history. Audit alerts appear in a non-intrusive right-hand panel, preserving the primary document view.

\subsection{Three-Stage Audit Pipeline}
The central technical contribution of Co-Refine is a three-stage audit pipeline that triggers automatically on every coding decision (and on sibling re-audits for overlapping segments) without interrupting the researcher’s workflow. Results are delivered via WebSocket events.

\paragraph{Stage 1: Deterministic Scoring}
An Azure OpenAI embedding model (\texttt{text-embedding-3-small}) maps each new segment \(\mathbf{e}_s \in \mathbb{R}^d\) and stores it in a per-user ChromaDB collection. For a code \(c\) with \(n\) prior segments, the centroid \(\boldsymbol{\mu}_c\) is the \(\ell_2\)-normalised mean vector:
\begin{equation}
\boldsymbol{\mu}_c = \frac{\bar{\mathbf{e}}}{\|\bar{\mathbf{e}}\|_2}, \qquad
\bar{\mathbf{e}} = \frac{1}{n}\sum_{i=1}^n \mathbf{e}_i
\label{eq:centroid}
\end{equation}
Centroid similarity is the cosine similarity:
\begin{equation}
\text{sim}(\mathbf{e}_s, \boldsymbol{\mu}_c) = \mathbf{e}_s \cdot \boldsymbol{\mu}_c
\label{eq:cosine}
\end{equation}
(Values in \([0.85, 1.0]\) = strong consistency; \([0.65, 0.85)\) = moderate; \(<0.65\) = flagged.)

Temporal drift is computed from the five oldest versus five newest segments when \(\geq10\) segments exist:
\begin{equation}
\delta_{\text{drift}} = 1 - \frac{\boldsymbol{\mu}_{\text{old}} \cdot \boldsymbol{\mu}_{\text{new}}}{\|\boldsymbol{\mu}_{\text{old}}\|_2 \|\boldsymbol{\mu}_{\text{new}}\|_2}
\label{eq:drift}
\end{equation}
Code overlap between any pair of codes is likewise measured via centroid cosine similarity (pairs \(>0.85\) are flagged for possible merge).

Cold-start handling: when a code has fewer than \(\tau_{\min}=3\) segments, its user-provided definition is embedded and used as a \emph{pseudo-centroid}; the UI and LLM context clearly flag this condition.

\paragraph{Stage 2: Grounded LLM Audit}
Stage-1 metrics, surrounding context, prior segments (sampled via recency + Maximal Marginal Relevance), and the latest Stage-3 code reflection are passed to a reasoning model (GPT-5.2). The model returns a structured JSON object whose \texttt{consistency\_score} is strictly constrained to \(\pm0.15\) of the Stage-1 centroid similarity; any deviation must be explicitly justified. The output also includes intent alignment, severity, headline, plain-language finding, drift warning, action suggestion, and alternative codes.

\paragraph{Stage 3: Code Reflection}
Triggered when a code reaches the auto-analysis threshold (default 3 segments) and every 3 additions thereafter, a reasoning model synthesises an evolving definition, theoretical lens, and derivation trace from a diverse sample of up to 30 segments. This reflection is injected into all future Stage-2 prompts, creating a deepening feedback loop.

\subsection{System Architecture and Implementation}
\begin{figure}[t]
    \centering
    \includegraphics[width=\linewidth]{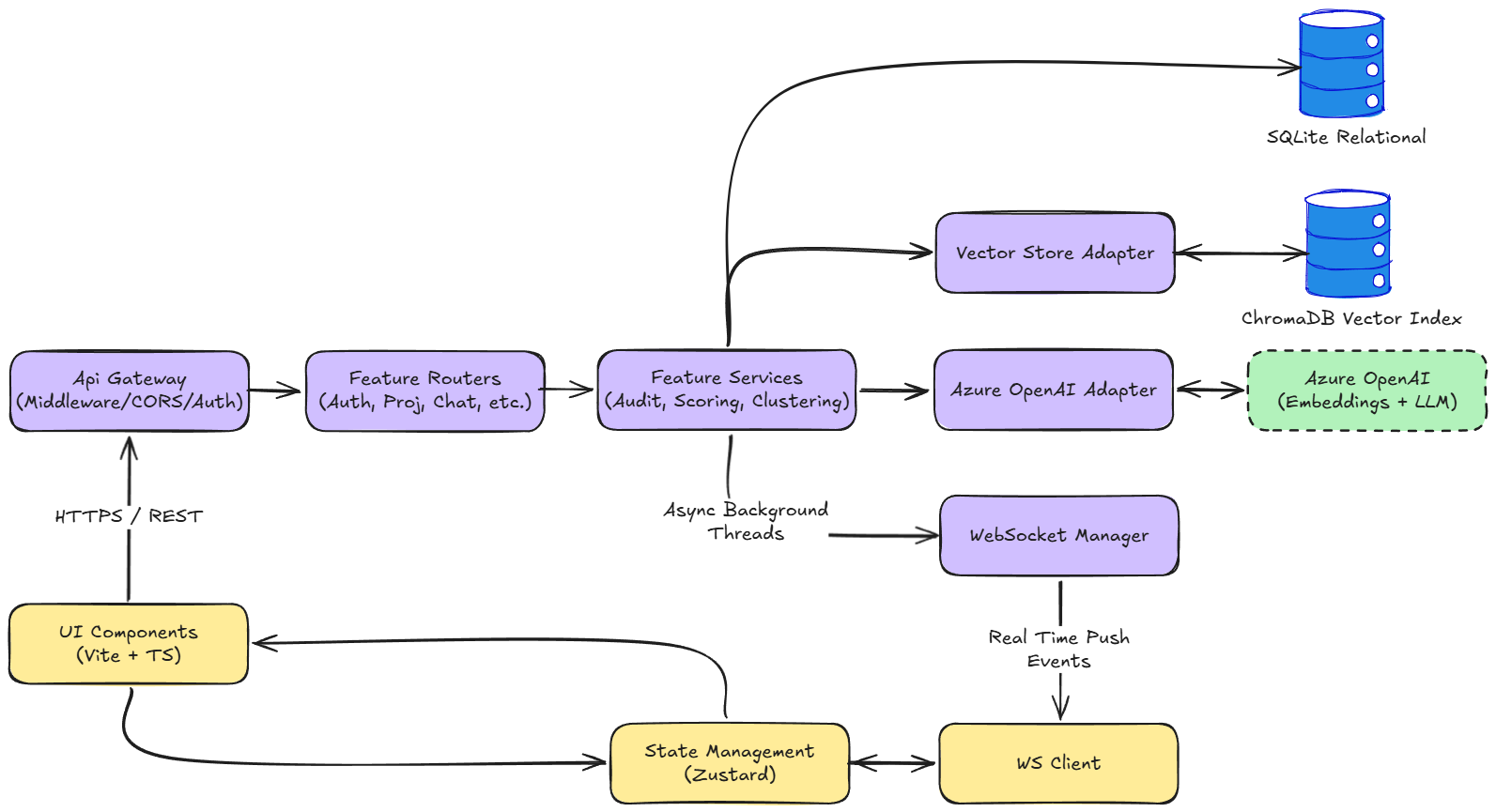}
    \caption{High-level architecture of Co-Refine.}
    \label{fig:arch-overview}
\end{figure}

Co-Refine follows a client-server architecture (Figure~\ref{fig:arch-overview}). The React 19 frontend (TypeScript, Zustand slice pattern, Tailwind CSS) communicates with a FastAPI backend via REST and authenticated WebSockets. Persistence uses SQLite (SQLAlchemy ORM) for relational data and per-user ChromaDB collections for vectors. LLM calls are routed through Azure OpenAI with two deployment tiers: a fast model for chat and facet labelling, and a reasoning model for audit and reflection. All primary keys are UUIDs; authentication tokens are stored in \texttt{sessionStorage} for security.

Key design decisions include append-only \texttt{ConsistencyScore} records (for trend analysis and audit trails), per-user vector collections, and deterministic grounding of LLM outputs. Background tasks use dedicated database sessions to ensure thread safety.

\subsection{User Interface and Key Features}
\begin{figure}[t]
    \centering
    \includegraphics[scale=0.16]{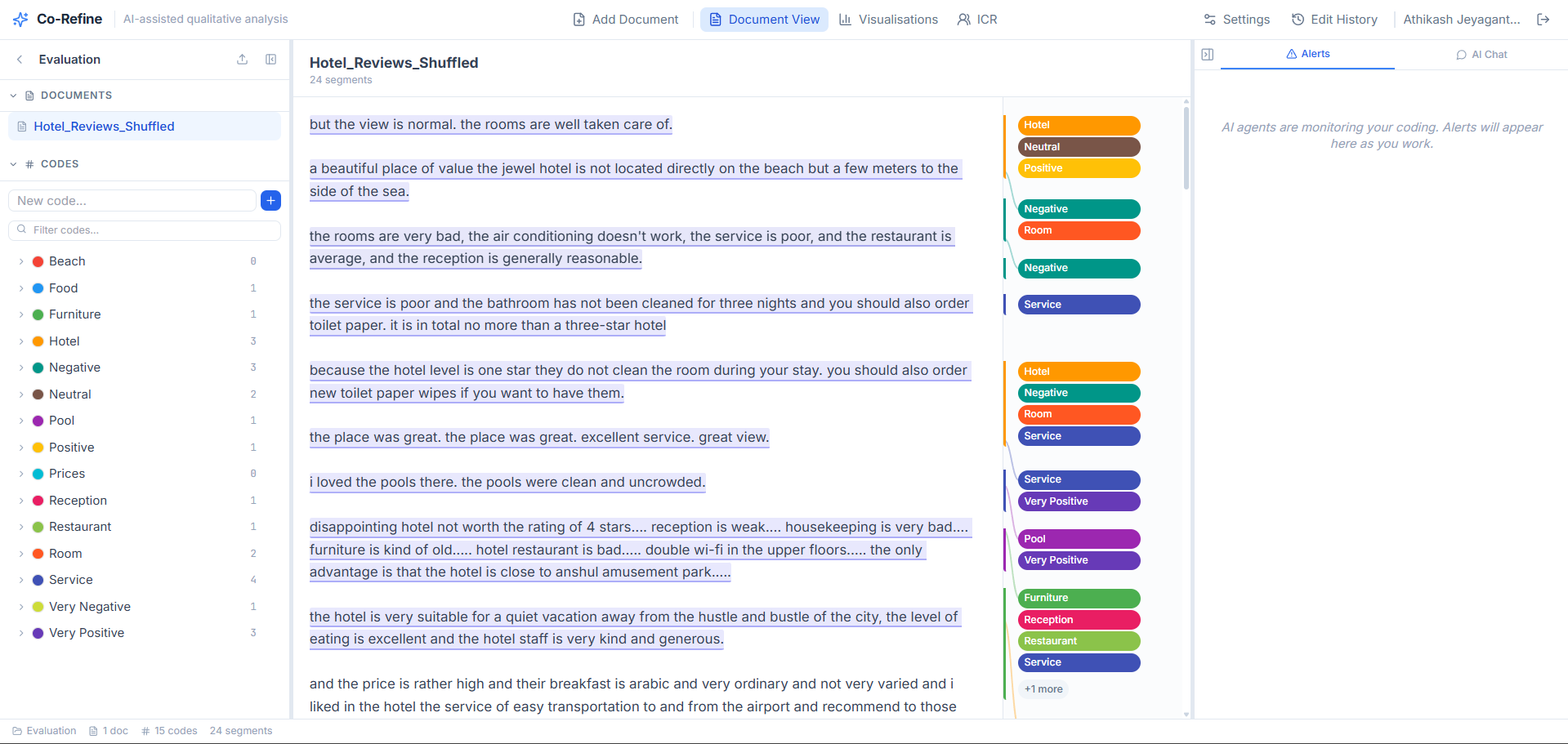}
    \caption{Document viewer with inline coding and audit alerts (right panel).}
    \label{fig:impl-doc-viewer}
\end{figure}

The interface follows a three-panel layout that mirrors the natural qualitative workflow (Figure~\ref{fig:impl-doc-viewer}). The left panel lists documents and the colour-coded codebook. The centre panel displays the full document with inline highlighting of coded segments. The right panel contains two tabs: real-time audit alerts and an AI chat assistant.

\begin{figure}[t]
    \centering
    \includegraphics[scale=0.5]{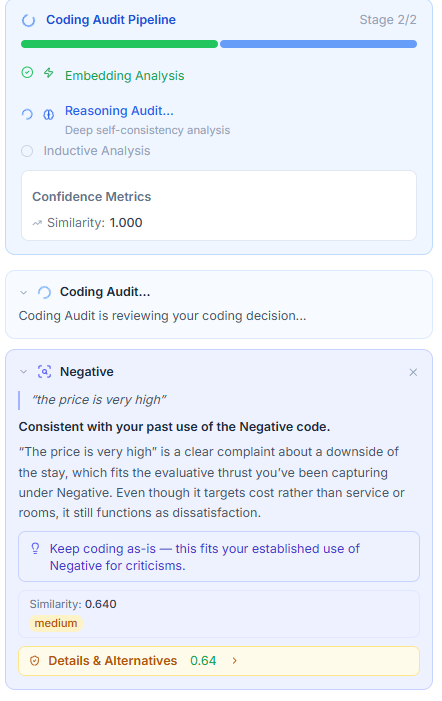}
    \caption{Real-time audit alert showing grounded LLM feedback.}
    \label{fig:impl-audit}
\end{figure}

Text selection triggers a popover lets researchers preview the span, apply one or more codes, and immediately receive an audit card (Figure~\ref{fig:impl-audit}) showing severity, deterministic scores, plain-language explanation, and an actionable suggestion. Researchers can dismiss or act on any alert.

Supporting features include:
\begin{itemize}
    \item An interactive visualisations dashboard with overview metrics, facet explorer (t-SNE + KMeans clustering), consistency timeline, code-overlap heatmap, and co-occurrence matrix.
    \item Collaborative ICR panel computing Cohen’s \(\kappa\), Fleiss’ \(\kappa\), and Krippendorff’s \(\alpha\), with disagreement classification and LLM-assisted resolution.
    \item Immutable edit history and project-level settings for threshold configuration.
\end{itemize}

The system was evaluated through unit, integration, and end-to-end tests (Playwright), accessibility auditing (axe-core, WCAG 2.1 AA), and a formative user study. Full details appear in Section~\ref{sec:eval}.

\section{Evaluation}
\label{sec:eval}

We evaluated Co-Refine through automated software testing, pipeline correctness checks on synthetic data, and a formative user study with five participants. The evaluation focused on technical reliability, usability, and perceived value for maintaining intra-coder consistency.

\subsection{Testing Strategy}

Automated tests were implemented at three levels. Unit tests (pytest for backend, Vitest + React Testing Library for frontend) validated scoring functions, JSON parsing, component rendering, and pure utilities. Integration tests used FastAPI’s \texttt{TestClient} with an in-memory SQLite database and mocked LLM/ChromaDB services to verify authentication, cascading deletes, and error handling. End-to-end tests with Playwright exercised the full workflow in a Chromium browser: registration, project creation, document upload, text selection, code application, alert delivery, and tab switching. Accessibility was audited with axe-core at both component and page levels, targeting WCAG 2.1 AA compliance for keyboard navigation, ARIA labels, and colour contrast.

Pipeline correctness was assessed on a controlled dataset of consistent, boundary, and deliberately inconsistent segments. We verified that Stage-1 deterministic scores were accurate and that Stage-2 LLM consistency scores remained strictly within \(\pm 0.15\) of the centroid similarity.

\subsection{Experimental Setup}

All coding tasks and the user study used a resampled subset of the SemEval-2016 ABSA English hotel-review dataset~\cite{billa2024semeval2016resampled}. This mixed-sentiment corpus is accessible, domain-neutral, and allows participants from varied backgrounds to focus on tool interaction rather than subject-matter expertise.

A formative user study was conducted with five participants (P1--P5) representing a range of expertise (Table~\ref{tab:user-background}). After a brief tutorial, each completed a 30--45 minute coding task on the dataset. Sessions ended with a modified 9-item System Usability Scale (SUS) questionnaire (scaled to 100) and a semi-structured interview. Thematic analysis of the transcripts identified perceptions of trust, cognitive load, and workflow integration.

\begin{table}[t]
\centering
\caption{Participant backgrounds in the formative user study.}
\label{tab:user-background}
\footnotesize
\setlength{\tabcolsep}{0pt} 
\begin{tabular*}{\columnwidth}{@{\extracolsep{\fill}} ll c c l l c}
\toprule
\textbf{ID} & 
\textbf{Role} & 
\makecell[b]{\textbf{Qual.}\\ \textbf{Res.}} & 
\makecell[b]{\textbf{Coding}\\ \textbf{(hrs)}} & 
\makecell[b]{\textbf{Prev.}\\ \textbf{Tools}} & 
\makecell[b]{\textbf{AI}\\ \textbf{Comfort}} & 
\makecell[b]{\textbf{Prior AI}\\ \textbf{Qual.}} \\
\midrule
P1 & PhD Stud.   & Yes & $>50$ & NVivo & Very comfort.   & Yes \\
P2 & UG Stud.    & No  & 0     & None  & Very comfort.   & No  \\
P3 & PhD Stud.   & Yes & 5--20 & None  & Somew. comfort. & Yes \\
P4 & UG Stud.    & Yes & $<5$   & None  & Very comfort.   & Yes \\
P5 & Acad. Staff & Yes & $>50$ & NVivo & Very uncomfort. & No  \\
\bottomrule
\end{tabular*}
\end{table}

\subsection{Results}

The audit pipeline performed as designed: deterministic Stage-1 scores were accurate, and the \(\pm0.15\) grounding constraint successfully eliminated contradictory LLM consistency scores observed in early prototypes. All generated alerts aligned with the expected classifications for consistent, boundary, and inconsistent segments.

SUS scores (Table~\ref{tab:sus}) averaged 77.77 (above the 68-point “above-average” threshold). Four of five participants rated the system “Excellent” ($\geq$80); the slightly lower score from P5 reflected initial intimidation when encountering active AI alerts. Thematic analysis of interviews revealed three recurring themes:

\begin{itemize}
    \item \textbf{Trust through grounding.} Participants valued that alerts were mathematically anchored in their own coding history rather than generic LLM opinions. P1 described the system as a “secondary triangulator” that provided high-level error checking (e.g., misapplying “perception” for “reception”).
    \item \textbf{Reflective friction without overload.} Alerts forced useful pauses (“cannot be ignored”) yet remained dismissible. The right-hand panel and red-highlighted segments allowed researchers to continue coding and review only when ready.
    \item \textbf{Desire for deeper theoretical grounding.} While the pipeline excelled at intra-code consistency, P1 noted it currently lacks explicit connection to broader theoretical frameworks; participants suggested future extensions such as uploading research papers to contextualise suggestions.
\end{itemize}

\begin{table}[t]
\centering
\caption{Scaled SUS scores (mean = 77.77).}
\label{tab:sus}
\small
\begin{tabular}{lcc}
\toprule
\textbf{ID} & \textbf{Adjusted Raw (max 36)} & \textbf{Scaled SUS (max 100)} \\
\midrule
P1 & 29 & 80.55 \\
P2 & 30 & 83.33 \\
P3 & 31 & 86.11 \\
P4 & 27 & 75.00 \\
P5 & 23 & 63.88 \\
\bottomrule
\end{tabular}
\end{table}

Participants also appreciated the facet explorer for surfacing sub-themes and the overlap heatmap for guiding code merges, confirming that the visualisations supported reflective practice.

\section{Discussion}

Co-Refine demonstrates that real-time, grounded AI assistance can meaningfully improve intra-coder consistency without automating interpretation. The \(\pm0.15\) deterministic grounding mechanism proved effective at constraining LLM outputs, addressing a core concern in prior LLM-assisted qualitative tools: hallucinated or ungrounded scores.

\subsection{Interpretation of Results}

Fully automated approaches such as LOGOS~\cite{pi2026logosllmdrivenendtoendgrounded} aim to replace human coding for scalability, achieving 80--88\% alignment with expert schemes. While impressive, such systems remove the researcher’s adaptive, bottom-up interpretive process—an essential element of reflexive thematic analysis (Braun \& Clarke). Co-Refine takes the opposite stance: it uses computational power to \emph{support} rather than replace human judgment. By continuously auditing every coding decision against the researcher’s own history, the system surfaces temporal drift and semantic overlap at the moment they occur, turning consistency maintenance from a burdensome post-hoc task into an integrated reflective practice.

This design aligns particularly well with later phases of reflexive thematic analysis (phases 4--5: reviewing and defining themes). The deterministic metrics and code-reflection stage force explicit comparison between intended and actual code usage, addressing the very pain points participants described in requirements elicitation.

\subsection{Comparison with Existing Tools}

Table~\ref{tab:tool-comparison} contrasts Co-Refine with representative CAQDAS and AI-assisted tools. No prior system combines real-time deterministic scoring, constrained LLM feedback, drift detection, and integrated ICR resolution workflows.

\begin{table}[t]
\centering
\caption{Comparison of Co-Refine with related qualitative analysis tools.}
\label{tab:tool-comparison}
\small
\setlength{\tabcolsep}{3pt} 
\begin{tabularx}{\columnwidth}{lccccc}
\toprule
\textbf{Qualitative Tool} & 
\makecell[b]{\textbf{Real-time} \\ \textbf{audit}} & 
\makecell[b]{\textbf{Det.} \\ \textbf{scoring}} & 
\makecell[b]{\textbf{LLM-} \\ \textbf{grounded}} & 
\makecell[b]{\textbf{ICR+} \\ \textbf{resol.}} & 
\makecell[b]{\textbf{Drift} \\ \textbf{detect.}} \\
\midrule
NVivo / ATLAS.ti & \texttimes & \texttimes & \texttimes & Partial & \texttimes \\
CollabCoder~\cite{gao2024collabcoder} & \texttimes & \texttimes & \checkmark & \checkmark & \texttimes \\
CoAIcoder~\cite{gao2023coaicode} & \texttimes & \texttimes & \checkmark & \checkmark & \texttimes \\
PaTAT~\cite{gebreegziabher2023patat} & \texttimes & Partial & \texttimes & \texttimes & \texttimes \\
Cody~\cite{rietz2021cody} & \texttimes & \texttimes & \texttimes & \texttimes & \texttimes \\
\textbf{Co-Refine} & \checkmark & \checkmark & \checkmark & \checkmark & \checkmark \\
\bottomrule
\end{tabularx}
\end{table}

\subsection{Limitations}

Technical limitations include text-only support (audio/video require transcription) and dependence on the quality of the underlying embedding model; specialised or non-English data may yield less reliable similarity scores. Although the \(\pm0.15\) constraint greatly reduces score hallucination, narrative explanations can still occasionally mislead if researchers over-trust them without inspecting the deterministic evidence.

Methodologically, both the requirements elicitation (N=3) and formative user study (N=5) involved a convenience sample from UK institutions; findings may not generalise to non-English-speaking researchers or those with lower AI familiarity. Pipeline evaluation used synthetic test segments; performance on highly nuanced real-world qualitative data remains an open question. The SUS score of 77.77, while promising, comes from a formative rather than summative study.

\section{Conclusion and Future Work}

We presented Co-Refine, an AI-augmented qualitative coding platform that delivers real-time, grounded consistency feedback while preserving full researcher agency. Its novel three-stage audit pipeline combines deterministic embedding metrics (centroid similarity, temporal drift, overlap) with a constrained LLM layer and evolving code reflection, creating a deepening feedback loop that surfaces intra-coder drift as it occurs rather than after the fact.

\subsection{Contributions}

The primary contribution is the design and empirical demonstration of a hybrid deterministic-LLM audit pipeline that constrains LLM outputs to mathematical evidence, enabling trustworthy real-time assistance in qualitative workflows. Secondary contributions include:
\begin{itemize}
    \item Facet discovery via K-Means + silhouette-optimised clustering and t-SNE visualisation, helping researchers identify latent sub-themes within a single code.
    \item An integrated collaborative ICR panel with typed disagreement classification and LLM-assisted resolution, supported by an immutable audit trail.
    \item Design principles and interface patterns that prioritise researcher control, configurability, and plain-language feedback.
\end{itemize}

These elements address a long-standing gap in CAQDAS tools: the lack of continuous, actionable support for intra-coder consistency during active coding.

\subsection{Future Work}

Immediate extensions include support for audio/video via integrated transcription, local LLM deployment for sensitive data, and richer theoretical grounding (e.g., uploading key papers to contextualise prompts). At the collaborative level, we plan to add negotiation threads and code-proposal workflows.

Longer-term research opportunities include a controlled comparative study measuring temporal drift reduction when using Co-Refine versus traditional or fully automated tools, and longitudinal field deployments with larger, diverse researcher cohorts. Such studies would further validate the system’s impact on analytical rigour and the credibility of qualitative findings.

By embedding deterministic signals within an otherwise interpretive workflow, Co-Refine demonstrates how HCI can design AI systems that augment rather than replace human expertise—offering a promising path toward more reliable, reflexive, and scalable qualitative research.

\bibliographystyle{ACM-Reference-Format}
\bibliography{references}

\end{document}